# Quantum Optics and Quantum Electrodynamics of Strong Field Processes

**Marcelo F. Ciappina[1,2,3], Misha Yu. Ivanov[4], Maciej Lewenstein[5,6], Javier Rivera-Dean[5], Philipp Stammer[5,7] and Paraskevas Tzallas[8,9,10]**

[1] Department of Physics, Guangdong Technion - Israel Institute of Technology, 241 Daxue Road, Shantou, Guangdong, China, 515063
[2] Technion – Israel Institute of Technology, Haifa, 32000, Israel
[3] Guangdong Provincial Key Laboratory of Materials and Technologies for Energy Conversion, Guangdong Technion – Israel Institute of Technology, 241 Daxue Road, Shantou, Guangdong, China, 515063
[4] Max-Born-Institut für Nichtlineare Optik und Kurzzeitspektroskopie, im Forschungsverbund Berlin e.V. Max-Born-Straße 2 A, 12489 Berlin
[5] ICFO - Institute of Photonic Sciences, Castelldefels, 08860 Spain
[6] ICREA, Pg. Lluís Companys 23, 08010 Barcelona, Spain
[7] Atominstitut, Trchnische Universtität Wien, 1020 Vienna, Austria
[8] Foundation for Research and Technology-Hellas, Institute of Electronic Structure & Laser, GR-70013 Heraklion (Crete), Greece
[9] Center for Quantum Science and Technologies (FORTH-QuTech), GR-70013 Heraklion (Crete), Greece.
[10] ELI-ALPS, ELI-Hu Non-Profit Ltd., Dugonics tér 13, H-6720 Szeged, Hungary

E-mail: maciej.lewenstein@icfo.eu

## Status

In its beginnings, the physics of intense laser-matter interactions was the physics of multiphoton processes. The theory was reduced then to high-order perturbation theory, while treating matter and light in a quantum manner. With the advent of chirped pulse amplification developed by D. Strickland and G. Mourou, which enabled generation of ultra-intense, ultra-short, coherent laser pulses, the need for a quantum electrodynamics description of electromagnetic (EM) fields practically ceased to exist and lost relevance [1]. Contemporary attoscience (AS), and more generally ultrafast laser physics, awarded the Nobel Prize in 2023 to P. Agostini, F. Krausz, and A. L'Huillier, commonly uses the classical description of EM fields while keeping a fully quantum description of matter. The progress and successes of AS in the last 40 years have been spectacular, with an enormous amount of fascinating investigations in basic research and technology. Yet a central question remains: can ultrafast laser physics (ULP) continue to advance without reintroducing quantum electrodynamics (QED) and quantum optics (QO) into its description of light-matter interactions?

In fact, QO has advanced towards the generation, control, and application of non-classical light states—such as Fock, squeezed, cat, and entangled states—that underpin emerging quantum technologies. However, these developments have largely been confined to low-photon-number and weak-field regimes, rather than high-intensity laser physics.

Recent pioneering works have demonstrated that conditioning intense-field processes—especially high-harmonic generation (HHG) and above-threshold ionization (ATI)—on measurable observables allows the generation of high-photon-number non-classical light. Optical Schrödinger cat states [2-6], multimode squeezed fields [5, 7, 8], and light-matter entangled states can now be engineered using strong-field processes, moving quantum optics firmly into the high-intensity regime. Moreover, bright squeezed states with intensity enough to drive or perturb HHG [9-11] and strong-field ionization [12] have been investigated. These developments have led to the emergence of *Quantum Optics and Quantum Electrodynamics of Strong Field Processes* — unifying ULP and quantum information science.

## Current and future challenges

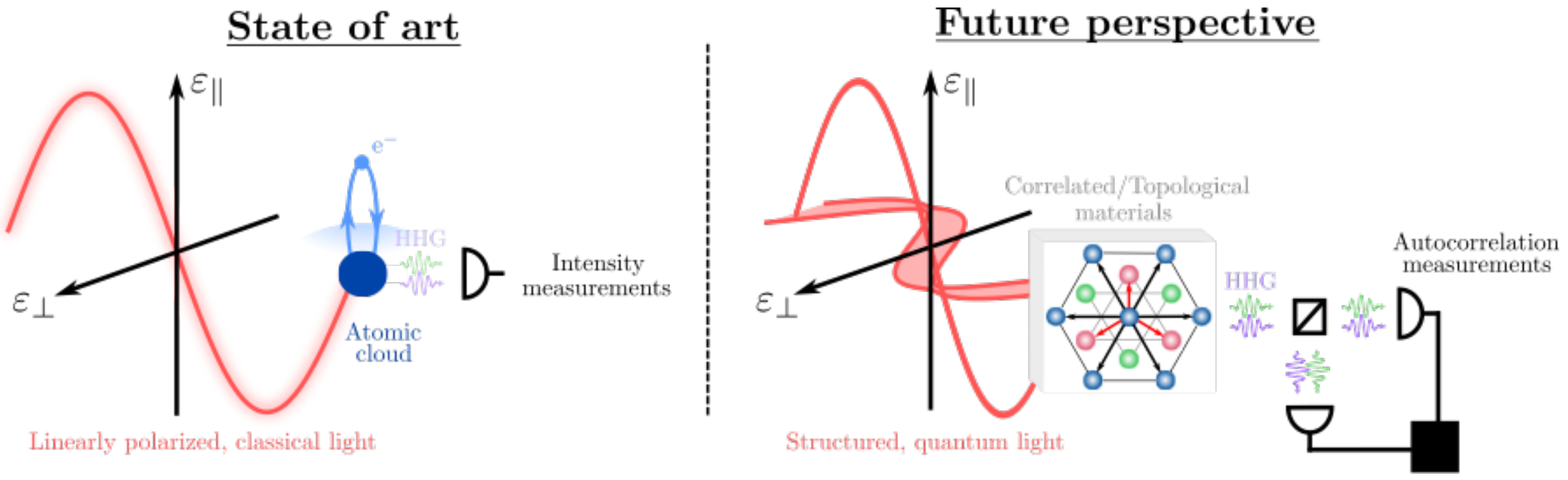

Figure 1. Intersection of quantum optics and strong-field physics Possible future direction to go beyond the state of the art in the intersection between quantum optics and strong-field physics. The scenarios considered recently include the use of high-photon-number structured quantum light, or quantum laser fields interacting with correlated or topological materials, with their outputs

There are four "most obvious" current and future challenges in this new area. They are all based on the necessary theoretical ingredient: development of quantum field-theoretical treatments of strong field physics [3].

**Challenge 1: Generation of massively quantum states of light.** New systematic methods should be established to produce large-photon-number entangled states, extending conditioning and post-selection techniques from atoms and molecules to correlated solids [4]. Generation of multimode squeezed and more exotic quantum states will be possible, going beyond the negligible depletion/excitation limit [8], or using HHG in resonant media [5]. Alternatively, one will use bright squeezed light [9, 10], or its mixture with conventional intense laser pulses [11], to generate high harmonics. Structured light will be used to study geometrical, chiral, and topological states in QED of attoscience, potentially enabling to probe and control many-body quantum systems at an unprecedented scale [4].

**Challenge 2: Generation of massively quantum states of light and matter.** One will explore entanglement between quantized light fields and electronic states in complex materials, focusing on ultrafast processes such as ATI, HHG, and rescattering. This will provide the first systematic framework for observing and exploiting light-matter entangled states in solids, using similar approaches as mentioned above. Earlier work has shown that conditioning on ATI events or on distinct HHG recombination paths in molecules and simple solids can generate light-matter entanglement. Recent advances [13] concerning the reconstruction of the photoelectron density matrix provide a new testing ground. These studies have revealed how classical and quantum noise reduce purity, but the open question remains: how does photon entanglement influence the reconstructed density matrix, and can experimental data unambiguously signal the presence of underlying quantum correlations? This connects QED of attoscience to the broader framework of multi-fragment "Zerfall" processes.

**Challenge 3: Characterization and exploitation of the generated states.** One will have to develop ultrafast quantum-optical methods to verify and quantify entanglement, providing new tools for nonlinear optics, precision metrology, and quantum information. This will open pathways to applying attosecond-scale entanglement across multiple fields. For the first steps toward applications in metrology and nonlinear optics, see [6, 14].

**Challenge 4: Simulation of QED of attoscience with quantum platforms.** One may be able to design and build quantum simulators based on cold atoms and ions, enabling controlled studies of strong-field dynamics such as quantum trap shaking [15], and generalized Kramers-Henneberger transformations. This will establish a new route to test and extend attoscience concepts under fully tunable conditions.

**Advances in science and technology to meet challenges**

To overcome the outlined challenges, recent advances in science and technology provide critical momentum for QED-QO-AS.

1. **Theoretical Progress:** Fully quantized descriptions of strong-field processes, while still being actively developed, are becoming feasible, thanks to both analytical and numerical advances. Models using quantum field operators and conditioning on specific observables (e.g. harmonics) allow us to derive entangled light-matter final states. QFT-based energy conservation has been introduced in post-selection schemes, moving beyond phenomenological assumptions. Novel hybrid approaches combine semiclassical dynamics for matter with quantum descriptions of light, balancing computational cost and physical insight.
2. **Experimental Techniques:** Attosecond metrology now includes schemes capable of characterizing photon statistics, coherence functions, and even probing Wigner functions of the emitted light. New platforms, such as ultrafast streaking and pump-probe configurations, enable time-resolved studies of quantum correlations and their possible impact on the quantum properties of the generated light [16]. Experimental demonstrations of non-classical states, such as optical Schrödinger cats and squeezed harmonics, have confirmed theoretical predictions and shown control over the quantum degrees of freedom.
3. **Structured and Quantum Light Sources:** The use of structured driving fields—including LG beams, Bessel beams, and polarization-shaped pulses—should allow new forms of MQS generation. Additionally, bright squeezed light and entangled photon sources from optical parametric amplifiers are being integrated with strong-field platforms. These sources open possibilities for high-fidelity quantum control and for going beyond the low-depletion regime.
4. **Material Platforms:** Advances in material science—2D materials, correlated systems, and topological insulators—offer novel testbeds for QO. HHG in these systems reveals phase-sensitive emission, quantum pathways, and evidence of entanglement in radiation. For example, in strongly-correlated materials [17, 18] and topological systems [19, 20], harmonic emission can encode correlations and quantum phase information.
5. **Quantum Simulators:** Cold atom and trapped ion simulators are now being used to mimic strong-field interactions using "quantum shaken traps" and Kramers-Henneberger transformations. These systems offer controlled platforms for benchmarking theory and testing QED-QO principles in clean, tunable settings. They also facilitate exploration of regimes not accessible in traditional solid-state systems.
6. **Measurement and Characterization Tools:** Advanced QO protocols—quantum tomography, homodyne detection, and entanglement witnessing—are being extended to the ultrafast domain. Multi-dimensional spectroscopy, quantum state reconstruction from interferometric data, and machine-learning-based inversion methods are being developed to diagnose MQS in real-time.

These technological innovations position the field to achieve reproducible, tunable, and scalable generation of high-photon-number quantum states for real-world applications.

**Concluding remarks**

Quantum electrodynamics and quantum optics of strong-field interactions are going towards a new frontier in ultrafast science—one where light and matter are treated as fully quantum entities, and where the quantum properties of intense laser fields are no longer ancillary, but central.

This emerging field offers a transformative perspective: strong-field processes are not merely tools for probing matter but can also be harnessed to engineer non-classical states of light and matter for quantum technologies. From the generation of attosecond-scale Schrödinger cat states to the entanglement of harmonic modes with conduction electrons, the interplay between intense fields and quantum coherence presents a wealth of opportunities.

Future breakthroughs will rely on increasing complexity: in materials (topological, chiral, correlated), in light (structured, squeezed, entangled), and in measurements (multi-mode quantum tomography, ultrafast entanglement diagnostics). The union of attosecond physics with quantum optics is no longer conceptual—it is being realized.

We envision a roadmap where the synergy between QED and strong-field physics enables scalable quantum light sources, new platforms for quantum simulation, and even ultrafast quantum processors. By deepening our understanding of quantum light-matter interactions at extreme scales, we pave the way for the next generation of quantum technologies.

## Acknowledgements

ICFO-QOT group acknowledges support from: European Research Council AdG NOQIA; MCIN/AEI (PGC2018-0910.13039/501100011033, CEX2019-000910-S/10.13039/501100011033, Plan National STAMEENA PID2022-139099NB, I00, project funded by MCIN/AEI/10.13039/501100011033 and by the "European Union NextGenerationEU/PRTR" (PRTR-C17.I1), FPI); QUANTERA DYNAMITE PCI2022-132919, QuantERA II Programme co-funded by European Union's Horizon 2020 program under Grant Agreement No 101017733; Ministry for Digital Transformation and of Civil Service of the Spanish Government through the QUANTUM ENIA project call - Quantum Spain project, and by the European Union through the Recovery, Transformation and Resilience Plan - NextGenerationEU within the framework of the Digital Spain 2026 Agenda; MICIU/AEI/10.13039/501100011033 and EU (PCI2025-163167); Fundació Cellex; Fundació Mir-Puig; Generalitat de Catalunya (European Social Fund FEDER and CERCA program; Barcelona Supercomputing Center MareNostrum (FI-2023-3-0024); Funded by the European Union (HORIZON-CL4-2022-QUANTUM-02-SGA PASQuanS2.1, 101113690, EU Horizon 2020 FET-OPEN OPTOlogic, Grant No 899794, QU-ATTO, 101168628), EU Horizon Europe Program (Grant No 101080086 NeQSTGrant Agreement 101080086 — NeQST). M. F. C. acknowledges support by the National Key Research and Development Program of China (Grant No.~2023YFA1407100), Guangdong Province Science and Technology Major Project (Future functional materials under extreme conditions - 2021B0301030005), the Guangdong Natural Science Foundation (General Program project No. 2023A1515010871) and the National Natural Science Foundation of China (Grant No. 12574092). M.I. acknowledges funding under DFG project number 545591821, IV 152/11-1. P.S. acknowledges funding from the European Union's Horizon 2020 research and innovation programe under the Marie Skłodowska-Curie grant agreement No 847517. P.T. acknowledges the Hellenic Foundation for Research and Innovation (HFRI) and the General Secretariat for Research and Technology (GSRT) under grant agreement CO2toO2 Nr.:015922, the European Union's HORIZON-MSCA-2023-DN-01 project QU-ATTO under the Marie Skłodowska-Curie grant agreement No 101168628 and ELI–ALPS. ELI–ALPS is supported by the EU and co-financed by the European Regional Development Fund (GINOP No. 2.3.6-15-2015-00001).

## References

1. Amini, K., Biegert, J., Calegari, F., Chacón, A., Ciappina, M. F., Dauphin, A., Efimov, D. K., de Morisson Faria, C. F., Giergiel, K., Gniewek, P. et al. (2019) Symphony on strong field approximation *Rep. Prog. Phys.*, **82**, 116001
2. Lewenstein, M., Ciappina, M. F., Pisanty, E., Rivera-Dean, J., Stammer, P., Th. Lamprou and Tzallas, P. (2021) Generation of optical Schrödinger cat states in intense laser–matter interactions. *Nat. Phys.*, **17**, 1104.
3. Stammer, P., Rivera-Dean, J., Maxwell, A., Th. Lamprou, Ordóñez, A., Ciappina, M. F., Tzallas, P. and Lewenstein, M. (2023) Quantum Electrodynamics of Intense Laser-Matter Interactions: A Tool for Quantum State Engineering *PRX Quantum*, **4**,010201.
4. Bhattacharya, U., Lamprou, Th., Maxwell, A.S., Ordóñez, A., Pisanty, E., Rivera-Dean, J., Stammer, P., Ciappina, M. F., Lewenstein, M., and Tzallas, P. (2023) Strong-laser-field physics, non-classical light states and quantum information science *Rep. Prog. Phys.,* **86**, 094401.
5. Yi, S., Klimkin, N. D., Brown, G. G., Smirnova, O., Patchkovskii, S., Babushkin, I. and Ivanov, M. (2025) Generation of Massively Entangled Bright States of Light during Harmonic Generation in Resonant Media *Phys. Rev. X*, **15**,011023.
6. Lamprou, T., Rivera-Dean, J., Stammer, P., Lewenstein, M. and Tzallas, P. (2025) Nonlinear Optics Using Intense Optical Coherent State Superpositions *Phys. Rev. Lett.*, **134**, 013601.
7. Theidel, D., Cotte, V., Sondenheimer, R., Shiriaeva, V., Froidevaux, M., Severin, V., Mosel, P., Merdji, H., Larue, A., Fröhlich, S. et al. (2024) Evidence of the quantum-optical nature of high-harmonic generation *PRX Quantum*, **5**,040319.
8. Stammer, P., Rivera-Dean, J., Maxwell, A.S., Lamprou, Th., Argüello-Luengo, J., Tzallas, P., Ciappina, M.F., and Lewenstein, M. (2024) Entanglement and Squeezing of the Optical Field Modes in High Harmonic Generation *Phys. Rev. Lett.* **132**, 143603.
9. Gorlach, A., Tzur, M.E., Birk, M., Krüger, M., Rivera, N., Cohen, O. and Kaminer, I., (2023). High-harmonic generation driven by quantum light, *Nat. Phys.*, **19**, 1689.
10. Rasputnyi, A., Chen, Z., Birk, M., Cohen, O., Kaminer, I., Krüger, M., Seletskiy, D., Chekhova, M. and Tani, F. (2024). High-harmonic

generation by a bright squeezed vacuum. *Nat. Phys.*, **20**, 1960–1965.

11. Lemieux, S., Jalil, S.A., Purschke, D.N., Boroumand, N., Hammond, T.J., Villeneuve, D., Naumov, A., Brabec, T. and Vampa, G., (2025), Photon bunching in high-harmonic emission controlled by quantum light, *Nat. Photon.*, **19**, 767.
12. Heimerl, J., Mikhaylov, A., Meier, S., Höllerer, H., Kaminer, I., Checkhova, M., and Hommelhoff, P. (2024) Multiphoton electron emission with non-classical light, *Nat. Phys.*, **20**, 945.
13. Laurell, H., Luo, S., Weissenbilder, R., Ammitzböll, M., Ahmed, S., Söderberg, H., Petersson, C. L. M., Poulain, V., Guo, C., Dittel, C. et al. (2025) Measuring the quantum state of photoelectrons *Nat. Photon.*, **19**, 352.
14. Stammer, P., Martos, T. F., Lewenstein, M. and Rajchel-Mieldzioć, G. (2024) Metrological robustness of high photon number optical cat states *Quant. Sci. Tech.*, **9**, 045047.
15. Argüello-Luengo, J., Rivera-Dean, J., Stammer, P., Maxwell, A., Weld, D., Ciappina, M., and Lewenstein, M. (2024) Analog Simulation of High-Harmonic Generation in Atoms *PRX Quantum*, **5**, 010328.
16. Lange, C.S., Hansen, T. and Madsen, L.B., (2024), Electron-correlation-induced nonclassicality of light from high-order harmonic generation. *Phys. Rev. A,* **109**, 033110.
17. Alcalà, J., Bhattacharya, U., Biegert, J., Ciappina, M., Elu, U., Graß, T., Grochowski, P.T., Lewenstein, M., Palau, A., Sidiropoulos, T.P. and Steinle, T., (2022), High-harmonic spectroscopy of quantum phase transitions in a high-Tc superconductor, *PNAS*, **119**, e2207766119.
18. Valmispild, V.N., Gorelov, E., Eckstein, M., Lichtenstein, A.I., Aoki, H., Katsnelson, M.I., Ivanov, M.Y. and Smirnova, O. (2024) Sub-cycle multidimensional spectroscopy of strongly correlated materials, *Nat. Photon.*, **18**, 432
19. Schmid, C.P., Weigl, L., Grössing, P., Junk, V., Gorini, C., Schlauderer, S., Ito, S., Meierhofer, M., Hofmann, N., Afanasiev, D., Crewse, J, Koch, K., Tereshenko, O., Güdde, J., Evers, F., Wilhelm, J., Richter, K., Höfer, U., and Huber, R., Tunable non-integer high-harmonic generation in a topological insulator, (2021), *Nature*, **593**, 385.
20. Neufeld, O., Tancogne-Dejean, N., Hübener, H., De Giovannini, U. and Rubio, A., (2023), Are there universal signatures of topological phases in high-harmonic generation? Probably not. *Phys. Rev. X*, ***13***, 031011.